\documentclass[final,3p,times,twocolumn]{elsarticle}

\journal{******}

\usepackage{amsmath}
\usepackage{amssymb}
\usepackage{makeidx}
\usepackage{gensymb}
\usepackage{amsfonts}
\usepackage[usenames,dvipsnames]{pstricks}
\usepackage{pst-grad} 
\usepackage{graphicx}
\usepackage{caption}
\usepackage{color}
\usepackage{allpurpose}

\begin{document}

\begin{frontmatter}



\title{Constraining Chaplygin models using diffuse supernova neutrino background}

\author[1]{Nan Yang}
\ead{nanyang27@gmail.com}
\author[2]{Junji Jia}
\ead{junjijia@whu.edu.cn}
\author[3,2]{Xionghui Liu}
\ead{liuxionghui@whu.edu.cn}
\author[4,5]{Hongbao Zhang}
\ead{hzhang@vub.ac.be}
\address[1]{Glyn O. Phillips Hydrocolloid Research Centre,  Hubei University of Technology, Wuhan 430068, China}
\address[2]{Center for Astrophysics \& Center for Theoretical Physics, School of Physics and Technology, Wuhan University, Wuhan, 430072, China}
\address[3]{School of Physics Science \& Technology, Lingnan Normal University, Zhanjiang, Guangdong 524048, China}
\address[4]{Department of Physics, Beijing Normal University, Beijing 100875, China}
\address[5]{Theoretische Natuurkunde, Vrije Universiteit Brussel and The International Solvay Institutes,
Pleinlaan 2, B-1050 Brussels, Belgium}

\begin{abstract}

In this work, we examine the possibility of using the diffuse supernova neutrino background (DSNB) to test the Chaplygin gas (CG) models of the Universe. With a typical supernova rate $R_{\mathrm{SN}}(z)$ and supernova neutrino spectrum $\dd N(E_\nu)/\dd E_\nu$, the DSNB flux spectrum $n(E_\nu)$ in three categories of CG models, the generalized CG (GCG), modified CG (MCG) and extended CG (ECG) models, are studied. It is found that generally the flux spectra take a form similar to a Fermi-Dirac distribution with a peak centered around 3.80-3.97 MeV. The spectrum shape and peak positions are primarily determined by $R_{\mathrm{SN}}(z)$ and  $\dd N(E_\nu)/\dd E_\nu$ and only slightly affected by the CG models. However, the height of the spectra in each category of the CG models can vary dramatically for different models, with variances of 13.2\%, 23.6\% and 14.9\% for GCG, MCG and ECG categories respectively. The averaged total flux in each category are also different, with the ECG model average 10.0\% and 12.7\% higher than that of the GCG and MCG models. These suggest that the DSNB flux spectrum height and total flux can be used to constrain the CG model parameters, and if the measured to a sub-10\% accuracy, might be used to rule out some models.

\end{abstract}


\end{frontmatter}


\section{Introduction}
\label{secint}

The cosmological observation data, including type Ia supernova (SN Ia) \cite{Riess:1998cb,Perlmutter:1998np, Kowalski:2008ez, Amanullah:2010vv,Suzuki:2011hu}, cosmic microwave background (CMB) \cite{Spergel:2003cb, Komatsu:2008hk, Komatsu:2010fb, cmbdataweb}, baryon acoustic oscillations (BAO) \cite{Eisenstein:2005su,Abazajian:2008wr,Percival:2009xn,Beutler:2011hx,Blake:2011en,Hinshaw:2012aka}, the observational Hubble data (OHD) \cite{Simon:2004tf},  implies the late-time cosmic acceleration. The most popular explanation is that about 70\% of the energy density of the Universe today are dark energy. One of the simplest form of this energy is the cosmological constant $\Lambda$ which is used in the $\Lambda$-cold dark matter model ($\Lambda$CDM) of the Universe. This constant is characterized by the equation of state (EoS) index $w=-1$, which is very consistent with the data constrained value $w=-1.03\pm 0.03$ \cite{Aghanim:2018eyx}.

However, the theoretical origin of $\Lambda$ has not been understood yet. If it was the vacuum energy associated with particle physics then it is hard to explain the huge gap between the theoretical and observed values \cite{Weinberg:1988cp,Carroll:2000fy}. Thus alternative theories attempting to explain the late-time acceleration without using constant dark matter are intensively studied. Broadly, these alternatives can be classified into two categories. The first category  is usually inside the framework of general relativity and based on a particular form of matter/field, such as quintessence \cite{Zlatev:1998tr,Amendola:1999er}, k-essence \cite{ArmendarizPicon:2000dh,ArmendarizPicon:2000ah,Scherrer:2004au}
 and Chaplygin gas (CG) \cite{Bento:2002ps, Gorini:2002kf, Bento:2002yx,Barreiro:2008pn, Lu:2008zzb,Thakur:2009jg,Lu:2010zzj,Xu:2012ca,Paul:2013sha,Khurshudyan:2014ewa, Pourhassan:2014ika, Paul:2014kza,Zhu:2015pta,Sharov:2015ifa,Thakur:2017syt,Paul:2017jrh}  or other fields/interactions. The second category however modifies either the general relativity or the homogenous and isotropic assumption of the universe, such as $f(R)$ theory \cite{Sotiriou:2008rp}, $f(T)$ theory \cite{Li:2010cg,Cai:2015emx}, inhomogeneous models \cite{Marra:2011ct, Bolejko:2011jc} and other models \cite{Bamba:2012cp}.

Among the first category, the CG models recently have gained much attention. It started from the basic CG model with simple EoS \cite{Gorini:2002kf}
\be P=-\frac{A}{\rho}\label{cgeos}\ee
where $A$ is a positive constant and then evolved to the generalized CG (GCG) model with EoS \cite{Bento:2002ps}
\be P=-\frac{A}{\rho^\alpha} \label{gcgeos}\ee
where $0<\alpha \leq 1$ in order for the Universe to enter the stage dominated by the cosmological constant. Currently, the simple CG model described by Eq. \refer{cgeos} has been ruled out completely \cite{Zhang:2004gc,Liao:2012gq}.
Later on, the EoS was further updated to the so-called modified CG (MCG) form
\be P=B\rho-\frac{A}{\rho^\alpha} ,\label{mcgeos}\ee
which is essentially a combination of an ordinary fluid obeying a linear barotropic EoS with the GCG fluid.
It is also possible to include fluid with constant, quadratic or even higher power barotropic EoS \cite{Khurshudyan:2014ewa, Pourhassan:2014ika, Paul:2017jrh} so that the model becomes the extended CG (ECG) with EoS
\be P=\sum_{i}B_i\rho^i-\frac{A}{\rho^\alpha} ,\label{ecgeos}\ee
where $B_i$ are also constants.
Clearly, setting proper constants to zero in Eq. \refer{ecgeos} will reduce it to \refer{mcgeos}, which further reduces to \refer{gcgeos} when $B=0$ and then to \refer{cgeos} when $\alpha$ is fixed to 1.

These various CG models, as well as the $\Lambda$CDM and some other dark matter alternative theories mentioned above, have been compared against observational data, including SN Ia, CMB, BAO, OHD and other data. It is generally found that all these models have some survival region in their respective parameter space. However, the question that which (kind of) CG models will be more correct than other CG models, or whether the CG models are more or less favored by the data, is not answered yet.

In this work, we discuss the possibility that these (x)CG models can be further constrained by the future diffuse supernova neutrino background (DSNB) measurement, or even discriminated if the DSNB measurement is accurate enough. When a star ends its life in the form of a supernova, almost its entire mass is released in the form of an enormous amount of neutrinos whose average energy is at the order $\cal{O}$(10) Mev level. Neutrinos from all supernovae in the history diffuse freely in the Universe and form the DSNB. The evolution of DSNB is influenced by how the Universe expands, just like the CMB or the cosmic neutrino background \cite{Jia:2008ti}, but at a much higher energy level. With the fast development of the neutrino observatory technologies and increase of observatory size in recent years, the observation of DSNB could be in the reach of a few upcoming neutrino experiments, such as  JUNO \cite{An:2015jdp}, Hyper Kamiokande \cite{Abe:2011ts} and some other experiments (see Ref. \cite{Lunardini:2010ab} for a review).

In what follows, we first briefly discuss  in section \ref{secsssr} the DSNB flux spectrum and the related core collapse (CC) supernova rate (SNR) and supernova neutrino spectrum (SNS). In section \ref{seccgsum} we summarize in details the aforementioned models and their corresponding cosmologies which are characterized by the Hubble parameters. We then use these inputs to compute the expected DSNB flux spectrum for each model in section \ref{seccomp}. It will be seen that for the parameters that are allowed by other data, the corresponding DSNB flux spectrum can differ by a large amount. Therefore if observed, the DSNB should provide a new way to severely constraint these models or even rule out some of them.

The idea of using DSNB to test cosmological models has previously been explored in Ref. \cite{Ono:2007zza} and  Ref. \cite{Barranco:2017lug}.
It was shown in the former that the GCG model has a total DSNB event that is 20\% larger than the $\Lambda$CDM and holographic dark energy models, and in the latter work that a bulk viscous matter-dominated universe can have larger total DSNB event than the $\Lambda$CDM and Logotropic universe models.
Our work differs from them in a few ways. Firstly, these previous works compared the total number of DSNB events of different kinds of cosmological models, whose analytical forms of the Hubble parameters are very different. While here we mainly test the ability of DSNB to distinguish sub-categories of the same broader kind of cosmological model -- the CG models. The difference between these sub-categories is much smaller than that between the cosmological models in the above references. Therefore our work illustrates that the DSNB can be used to discriminate models with more subtle differences. Secondly, in this work the effects of the variations in the star formation rate (SFR), CC SNR and SNS to the final DSNB spectrum are more thoroughly investigated than in previous works. We altered not only the parameter values (see Eqs. \eqref{stellarrate} and \eqref{abcdchange}, and temperature change in Eq. \eqref{snnspec}) but also the analytical forms of these quantities (compare Eqs. \eqref{stellarrate} and \eqref{stellarrate2}, and Eqs. \eqref{snnspec} and \eqref{snnspec2}) to compute and discuss in details the resultant DSNB spectra. In previous works, only some variation of the parameters in one such quantities (CC SNR in Ref. \cite{Ono:2007zza}) was computed. Thirdly, in our analysis, in addition to the total DNSB event which was the main quantity analyzed in previous works, we also analyzed the peak positions of the DNSB spectra and showed that they are primarily determined by the SNR and SNS but not much affected by the cosmological models.

\section{The DSNB flux density\label{secsssr}}

The spectrum of DSNB flux in the energy interval $[E_\nu,~E_\nu+\dd E_\nu]$ emitted in the redshift interval $[z,~z+\dd z]$ can be calculated using the ``line-of-sight'' integral method to be \cite{Ando:2004hc}
\be \dd n_\nu(E_\nu) = \left[ R_{\mathrm{SN}}(z)(1 + z)^3\right] \left[-\frac{\dd t}{\dd z}\dd z\right] \left[\frac{\dd N_\nu(E_\nu^\prime)}{\dd E_\nu^\prime} \dd E_\nu^\prime\right](1+z)^{-3}, \label{diffflux}\ee
where $E_\nu^\prime=(1+z)E_\nu$ is the neutrino energy at redshift $z$ which is now measured as $E_\nu$. $\displaystyle\frac{\dd N_\nu(E_\nu)}{\dd E_\nu}$ is the effective neutrino number spectrum for one supernova. $R_{\mathrm{SN}}(z)$ is the CC SNR at redshift $z$. The factors $(1+z)^{\pm 3}$ takes into account the effect of Universe expansion to the volume integral element. The term $\displaystyle -\frac{\dd t}{\dd z}$ is related to the Hubble parameter by
\be -\frac{\dd t}{\dd z}=\frac{1}{1+z}\frac{1}{H(z)}. \ee

The DSNB flux spectrum is obtained then by integrating Eq. \refer{diffflux} with respect to the redshift
\be n(E_\nu)=\int_0^\infty \left[ R_{\mathrm{SN}}(z) \frac{\dd N_\nu(E_\nu^\prime)}{\dd E_\nu^\prime}\right] \left[\frac{1}{H(z)}\right]\dd z . \label{dsnbflux}\ee
Therefore clearly the DSNB flux \refer{dsnbflux} depend on the cosmological model $H(z)$ very critically. Besides this, the CC SNR $R_{\mathrm{SN}}(z)$ and supernova spectrum $\displaystyle \frac{\dd N_\nu(E_\nu^\prime)}{\dd E_\nu^\prime}$ are the other two components under the integral.
For the CC SNR $R_{\mathrm{SN}}(z)$, it was thought that there were a problem of mismatch between it and the SFR \cite{Horiuchi:2011zz}. However recently a few  promising solution have been proposed including enhancing the initial mass function \cite{Mathews:2014qba}, revised star formation history \cite{Madau:2014bja} and change of the stellar mass ranges that end up as CC SNe \cite{Hidaka:2016zei}. After these solutions, it appears that the current measured CC SNR roughly reflect the true rate happens in the Universe. Moreover, since our concentration here is on the effect of cosmological models on the DSNB but not that of the SNR or SFRs, we will direct adopt the CC SNR in Refs.  \cite{Strolger:2015kra,Petrushevska:2016kie}, which is
\be R_{\mathrm{SN}}(z)=k\cdot h^2\cdot \psi(z) \label{snrate} \ee
where $k\simeq 0.0091~M_\odot^{-1}$ is the number of stars per unit mass that explode as CC SNe, $h$ is defined as in $H_0=100 h  \mbox{~km~s}^{-1}\mbox{Mpc}^{-1}$ and $\psi(z)$ is the SFR given by Ref.
\cite{Madau:2014bja,Petrushevska:2016kie}
\be \psi(z)= \frac{A(1+z)^C}{1+((1+z)/B)^D}~~[\mbox{M}_\odot \mbox{year}^{-1}\mbox{Mpc}^{-3}] \label{stellarrate}\ee
with $A=0.015,~B=2.9,~C=2.7$ and $D=5.6$.

For the single SNS, we only consider the anti-electron neutrino spectrum because the main  detection channel in the observatories with good DSNB capabilities such as JUNO or Hyper Kamiokande is the inverse beta decay  process $\bar{\nu}_e + P \to e^+ + n$ \cite{An:2015jdp,Abe:2011ts}.
This spectrum can be approximated by the pinched Fermi-Dirac spectrum \cite{Raffelt:1996wa,Beacom:2010kk}
\begin{align}
 \frac{\dd N_\nu(E_\nu^\prime)}{\dd E_\nu^\prime} =E_{\bar{\nu}_e,tot}\frac{120}{7{\pi}^4}
\frac{{E^\prime}_{\nu}^2}{T^4}\frac{1}{e^{E_{\nu}^\prime/T}+1}, \label{snnspec}
\end{align}
where we choose $E_{\bar{\nu}_e,tot}\sim5\times10^{52}$ erg to represent the typical energy radiated away by the neutrinos during supernova, $T=5$ Mev is the radiation temperature \cite{Beacom:2010kk}.

In principle, both of the SNR $R_{\mathrm{SN}}(z)$  and the neutrino spectrum $\displaystyle\frac{\dd N_\nu(E_\nu)}{\dd E_\nu}$ should depend not only on the redshift but also on other parameters such as the supernova progenitor mass. The form \eqref{stellarrate} and \eqref{snnspec} we used here are in this sense averaged over other factors. The parameters in them therefore carry their own uncertainties. However, as we will show in section \ref{secdis} these uncertainties will not lower the variation percentage of the resultant DSNB spectrum height or total flux among different CG models, and consequently  the constraining power of the DSNB to the CG models will not be lost.

\section{The (x)CG models \label{seccgsum}}

For the MCG model with EoS \refer{mcgeos}, solving the energy conservation equation one obtains the energy density of MCG given by
\begin{align} \rho_{\mathrm{MCG}}&=\rho_{0,\mathrm{MCG}}\left[A_S+\frac{1-A_S}{a^{3(1+\alpha)(1+B)}}\right]^{\frac{1}{1+\alpha}}\nonumber\\
&\equiv \rho_{0,\mathrm{MCG}} f(A_S,B,\alpha,a) \label{mcged},
\end{align}
where $a$ is the scale factor in the Friedmann-Robertson-Walker metric, $A_S=A/[(B+1)\rho_{0,\mathrm{MCG}}^{\alpha+1}]$ and $\rho_{0,\mathrm{MCG}}$ is a positive integration constant representing the MCG density today.
Further assuming that the Universe is filled with the MCG, regular baryonic matter and regular radiation components $\rho_{\mathrm{MCG}}$, $\rho_b$, $\rho_r$, and taking into account the curvature effective energy density $\rho_\kappa$ then the total density $\rho_t$ becomes
\be \rho_t=\rho_{\mathrm{MCG}}+\rho_b+\rho_r+\rho_\kappa.\ee

Making use of the Friedman equation, the Hubble parameter is obtained as
\begin{align}
 H_{\mathrm{MCG}}=&H_0\left\{(1-\Omega_{0,b}-\Omega_{0,r}-\Omega_{0,\kappa})f(A_S,B,\alpha,(1+z)^{-1}) \right.\nonumber\\
&\left.+\Omega_{0,b}(1+z)^3 +\Omega_{0,r}(1+z)^4 +\Omega_{0,\kappa} (1+z)^2\right\}^{1/2},
\label{mcghp}
\end{align}
where we have used $(1+z)^{-1}$ to replace $a$ and
$\Omega_{0,i}~ (i = b, r, k)$ are today's values of dimensionless energy densities of baryon, radiation and effective curvature respectively.
Most of the previous studies on the CG models (see Tables \ref{tbgcg} and \ref{tbecg}) assumed that $\Omega_{0,r}=\Omega_\kappa=0$ to simplify the cosmological models. For the DSNB flux calculation given by Eq. \refer{dsnbflux},  there is one more reason that they can be ignored: the neutrino spectrum \refer{snnspec} suppresses exponentially the contribution to the flux from small $a$ eras,  while for the large $a$ stage it is known that the radiation density as well as the effective curvature parameter are both small (Plank 2018 data yields a value of $\Omega_{0,\kappa}=0.001\pm 0.002$ \cite{Aghanim:2018eyx}).
Setting $B=0$ in Eq. \refer{mcghp} will produce the Hubble parameter for the GCG model
\begin{align}
H_{\mathrm{GCG}}=&H_0\left\{(1-\Omega_{0,b}-\Omega_{0,r}-\Omega_{0,\kappa})\right.\nonumber\\
&\times\left[A_S+(1-A_S)(1+z)^{3(1+\alpha)}\right]^{\frac{1}{1+\alpha}} \nonumber\\
&\left.+\Omega_{0,b}(1+z)^{3} +\Omega_{0,r}(1+z)^4 +\Omega_{0,\kappa} (1+z)^2\right\}^{\frac{1}{2}}. \label{gcghp}
\end{align}

While for the ECG models with EoS \refer{ecgeos}, the energy conservation equation in the general case are not analytically solvable to obtain solutions like Eq. \refer{mcged}. Only a handful of work are carried out with particularly chosen nonzero $B_i$ or $A$ in the EoS \refer{ecgeos} \cite{Kahya:2014fja,Sharov:2015ifa, Paul:2017jrh}. Here we will only concentrate on the following three cases considered by Ref. \cite{Paul:2017jrh} due to its explicit Hubble parameter formulas that can be directly used by our calculations. First  when the sum in Eq. \eqref{ecgeos} has only two nonzero term $i=1$ and $i=n$, then denoting $B_1=-A,~B_n=B$, the EoS becomes
\begin{align}
P=-A\rho+B\rho^n ,~\label{ecgeos1}\end{align}
and the corresponding Hubble parameter is solved to be
\begin{align}
H_{\mathrm{ECG1}}(z)=&H_0\left\{(1-\Omega_{0,b}-\Omega_{0,r}-\Omega_{0,\kappa})\right.\nonumber\\
&\times\left[A_S+(1-A_S)(1+z)^{3(A-1)(n-1)}\right]^{\frac{1}{1-n}}\nonumber\\
&\left.+\Omega_{0,b}(1+z)^{3} +\Omega_{0,r}(1+z)^4 +\Omega_{0,\kappa} (1+z)^2\right\}^{\frac{1}{2}}, \label{ecghp1}
\end{align}
in which $A_S, ~A$ are parameters converted from $A,~B$ and $n$.
When the EoS is given by
\begin{align}
P=B\lb \rho+\rho^2\rb -\frac{A}{\rho^{1/2}} ,~\label{ecgeos2}\end{align}
the Hubble parameter will be
\begin{align}
 H_{\mathrm{ECG2}}(z)=H_0\lsb1+\frac{1}{X^2}(1+z)^{[9(B+1)+15X^2B]/2}\rsb , \label{ecghp2}
\end{align}
where $X$ is some effective parameter related to $B$ in Eq. \eqref{ecgeos2}.
Finally when the EoS is
\begin{align}
P=B\lb \rho+\rho^2+\rho^3\rb-\frac{A}{\rho^{1/2}} ,~\label{ecgeos3}\end{align}
the Hubble parameter is
\begin{align}
 H_{\mathrm{ECG3}}(z)=H_0\lsb1+\frac{1}{Y^2}(1+z)^{[9(B+1)+15Y^2B+21Y^4B]/2}\rsb . \label{ecghp3}
\end{align}
Again, $Y$ here is an effective parameter related to $B$ in Eq. \eqref{ecgeos3}.

The $\Lambda$CDM model we used to compare to the above models is given by
\begin{align}\label{hlcdm}
H_{{\Lambda} { \mbox{\scriptsize CDM}}}&=H_0\left[\Omega_{m}(1+z)^3+\Omega_{\Lambda}\right]^{\frac{1}{2}},
\end{align}
where we take $\Omega_\Lambda=0.685$, $\Omega_m=0.315$ and  $H_0=67.4\mbox{~km~s}^{-1}\mbox{Mpc}^{-1}$ \cite{Aghanim:2018eyx}.

As mentioned earlier, these models were constrained using different combination of data sets and methods. Consequently, the obtained values of parameters are also slightly different even for the same kind of CG models. Here we summarize the values of the parameters in the GCG models (Table \ref{tbgcg}), MCG models (Table \ref{tbmcg}) and ECG models (Table \ref{tbecg}) according to the data sets that were used.

\begin{table*}[!htbp]
\begin{center}
\caption{The GCG model parameters in various references.  See Ref. \cite{Paul:2014kza} for the Growth and $\sigma_8$ data. The last two columns are the peak positions and heights in Fig. \ref{figcg}(a) with units MeV and MeV$^{-1}$cm$^{-2}$s$^{-1}$ respectively. \label{tbgcg}}
\begin{tabular}{l|l|l|l|l||l|l}
  \hline\hline
Model  \# & $\alpha$       & $A_S$       & Data set        & Ref. &P. P.&P. H.\\
  \hline
GCG1&  0.4    		&0.83 			& CMB+Quasar+SN   &\cite{Bento:2002yx}&3.955&0.794\\
GCG2&  0.1     		& 0.77        & CMB         		&\cite{Barreiro:2008pn}&3.945&0.737\\
GCG3&  0.00126	& 0.775       & BAO+CMB+SN      &\cite{Xu:2012qx}&3.926&0.755\\
GCG4&  0.141    	& 0.819      & \hspace{-2mm}\begin{tabular}{l} BAO+CMB+Growth\\
~~+OHD+SN+$\sigma_8$\end{tabular}     					& \cite{Paul:2014kza}&3.932&0.770\\
GCG5&  0.20    		& 0.77        & BAO+CMB+SN   &\cite{Zhu:2015pta}&3.950&0.701\\
GCG6&  -0.100  		& 0.738       & BAO+OHD+SN      	& \cite{Sharov:2015ifa}&3.915&0.767\\
GCG7&  -0.069  		& 0.753       & BAO+OHD+SN      	& \cite{Sharov:2015ifa}&3.921&0.768\\
GCG8&  0.023   		& 0.7720      & BAO+CMB+OHD+SN  &\cite{Thakur:2017syt}&3.934&0.760\\
GCG9&  0.4056       & 0.8846      & BAO+CMB+OHD     & \cite{Paul:2017jrh}&3.946&0.789\\
\hline\hline
\end{tabular}
\end{center}
\end{table*}

\begin{table*}[!htbp]
\begin{center}
\caption{The MCG model parameters in various references. See Ref. \cite{Lu:2010zzj, Allen:2007ue} for the CBF data and Ref. \cite{Paul:2013sha,Paul:2014kza} for the Growth and $\sigma_8$ data. The last two columns are the peak positions and heights in Fig. \ref{figcg}(b) with units MeV and MeV$^{-1}$cm$^{-2}$s$^{-1}$ respectively. \label{tbmcg}}
\begin{tabular}{l|l|l|l|l|l||l|l}
  \hline\hline
Model \# & $\alpha$ & $A_S$ & $B $ & Data set & Ref.&P. P.&P. H.\\
  \hline
MCG1&  1.724     & 0.822     		& -0.085    	& BAO+CMB+SN        & \cite{Lu:2008zzb}&3.964&0.638\\
MCG2&  0.11      & 0.8       			& 0.06      	& OHD               			& \cite{Thakur:2009jg}&3.973&0.702\\
MCG3&  0.1079    & 0.7788    		& 0.00189   	&\hspace{-2mm}\begin{tabular}{l} BAO+CBF+CMB\\
~~+OHD+SN\end{tabular}														& \cite{Lu:2010zzj}&3.951&0.741\\
MCG4&  0.000727  & 0.782     	& 0.000777  	& BAO+CMB+SN        & \cite{Xu:2012ca} &3.922&0.755\\
MCG5&  0.002     &0.769 			&0.008 & Growth+OHD+$\sigma_8$ & \cite{Paul:2013sha}&3.935&0.744\\
MCG6&  0.1905    & 0.8252 & 0.0046 & \hspace{-2mm}\begin{tabular}{l} BAO+CMB+Growth\\
~~+OHD+SN+$\sigma_8$\end{tabular} &									 \cite{Paul:2014kza}&3.943&0.762\\
MCG7&  0.538     & 0.714     		& -0.166    	& BAO+OHD+SN        & \cite{Sharov:2015ifa}&3.891&0.769\\
MCG8&  0.613     & 0.716     		& -0.176    	& BAO+OHD+SN        & \cite{Sharov:2015ifa}&3.890&0.769\\
MCG9&  0.5218    & 0.9036   		 & 0.0067    	& BAO+CMB+OHD       & \cite{Paul:2017jrh}&3.958&0.789\\
  \hline\hline
\end{tabular}
\end{center}
\end{table*}

\begin{table*}[!htbp]
\begin{center}
\caption{The ECG model parameters in three cases. The last two columns are the peak positions and heights in Fig. \ref{figcg}(c) with units MeV and MeV$^{-1}$cm$^{-2}$s$^{-1}$ respectively.\label{tbecg} }
\begin{tabular}{l|l|l|l||l|l}
  \hline\hline
Model \# & Parameters         & Data set        & Ref.    &P. P.&P. H.  \\
\hline
ECG1 & $A_S=0.9028,~A=0.0066, ~n=-0.5146$ & BAO+CMB+OHD       & \cite{Paul:2017jrh} &3.957&0.789\\
ECG2 & $B = -0.037662,~X =2.83633$   		  & BAO+CMB+OHD       & \cite{Paul:2017jrh}  &3.795&0.907   \\
ECG3 & $B = -0.00328,~Y =2.83641$   			  & BAO+CMB+OHD       & \cite{Paul:2017jrh}   &3.795&0.907  \\
\hline\hline
\end{tabular}
\end{center}
\end{table*}

\section{The DSNB flux in (x)CG cosmologies \label{seccomp}}
\begin{figure*}[!htbp]
\vspace{-0.5cm}
\begin{center}
\includegraphics[width=0.75\textwidth]{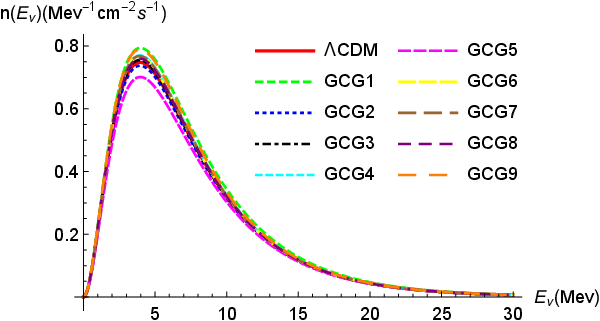}\\
(a)\\
\includegraphics[width=0.75\textwidth]{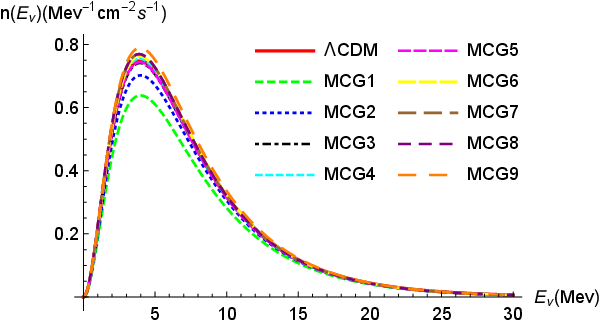}\\
(b)\\
\includegraphics[width=0.75\textwidth]{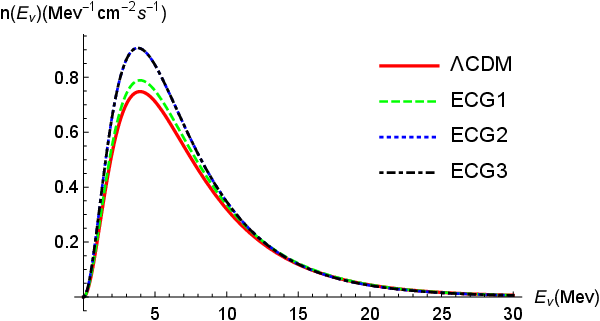}\\
(c)
\caption{The flux spectra in GCG models (a), MCG models (b) and ECG models (c). The flux spectrum in $\Lambda$CDM model is drawn in each subplot using red solid line. See last two columns in each of the Tables \ref{tbgcg}-\ref{tbecg} for the exact peak positions and heights. \label{figcg}}
\end{center}
\end{figure*}

With the CC SNR \eqref{snrate}, SFR \eqref{stellarrate}, SNS \refer{snnspec} as well as the cosmological models described by Eqs. \refer{mcghp}, \refer{gcghp}, \eqref{ecghp1}, \eqref{ecghp2}, \eqref{ecghp3} and \eqref{hlcdm}, we can now integrate Eq. \refer{dsnbflux} to find the DSNB flux for all models with parameters listed in Tables \ref{tbgcg}, \ref{tbmcg} and \ref{tbecg}.
The results are shown separately for the GCG, MCG and ECG models in Fig. \ref{figcg} (a),  (b) and (c).

It is seen that for all models with different values of parameters, the spectrum takes the shape similar to a Fermi-Dirac distribution, with a single peak centered around 3.80-3.97 MeV (the exact peak position of each model is given in Tables \ref{tbgcg}-\ref{tbecg}). This suggest that the shape of the spectrum, as well as the location of the peak, is not primarily determined by the cosmological model but other factors in the integral \eqref{dsnbflux}, namely the CC SNR and SNS. The cosmological models when limited by the range of the parameters in Tables \ref{tbgcg}-\ref{tbecg} however, still can weakly affect the peak positions.  This point is reflected by the fact that for the ECG models, especially the models ECG2 and ECG3 in Fig. \ref{figcg} (c), their peak positions are shifted to a slightly lower value comparing to those of the GCG and MCG models. A simple statistic analysis shows that the peak position average for the GCG models is 3.94$\pm$0.03 MeV. Average peak position of the  MCG models is found to be the same as the GCG models, but that of the ECG models is 3.85$\pm$0.09 MeV. Therefore the mean peak position of the ECG models considered here is about 2.3\% lower than that of the GCG or MCG models. Note that the neutrino observatory such as Juno can have an energy resolution of 3\%/$\sqrt{E/\mbox{MeV}}$ \cite{An:2015jdp} which is about 1.5\% around the peak energy here.

Unlike the spectrum shape or the peak positions, the heights of the peaks are strongly affected by the cosmological models (see Tables \ref{tbgcg}-\ref{tbecg} for the exact peak height for each model). From Fig. \ref{figcg} (a), it is seen that the maxima of the DSNB flux spectra in different GCG models can vary from 0.701 MeV$^{-1}$cm$^{-2}$s$^{-1}$ to 0.794 MeV$^{-1}$cm$^{-2}$s$^{-1}$, a difference about 13.2\%. For the MCG models, Fig. \ref{figcg} (b) shows that the highest and lowest of the peaks are respectively 0.638 MeV$^{-1}$cm$^{-2}$s$^{-1}$ and 0.789 MeV$^{-1}$cm$^{-2}$s$^{-1}$ with a difference of 23.6\%. The peak hight in different ECG models ranges  from 0.789 MeV$^{-1}$cm$^{-2}$s$^{-1}$ to 0.907 MeV$^{-1}$cm$^{-2}$s$^{-1}$, which corresponds to a difference of 14.9\%. It is clear that the DSNB flux spectrum height is very sensitive to the cosmological model, and this suggests that the future measured DSNB flux spectrum will be usable to constraint the Chaplygin gas models. If this data is measured to an accuracy of sub-10\% level, then it might even be used to tell which Chaplygin gas is more viable.

We also integrated the flux spectrum over energy to obtain the total flux for each model and then averaged over the same kind of models. It is found that for the GCG models considered in Table \ref{tbgcg}, MCG models in Table \ref{tbmcg} and ECG models in Table \ref{tbecg}, the average fluxes are respectively about  6.80 cm$^{-2}$s$^{-1}$, 6.63 cm$^{-2}$s$^{-1}$ and  7.47
cm$^{-2}$s$^{-1}$. The former two are close to that of the $\Lambda$CDM model given in Eq. \refer{hlcdm} which is 6.72 cm$^{-2}$s$^{-1}$ and the ECG model flux is about 10.0\%-12.7\% higher than them.
Therefore the total flux can also be used to constraint these models, even if the energy resolved spectrum is not easily obtained in the initial stages of the DSNB measurements.

\subsection*{Effects of model parameters}

In this subsection, we will discuss how the constraining power of the DSNB fluxes might be affected by the uncertainties of the model parameters. As mentioned earlier, these uncertainties mainly come from two sources, the SFR which determines the CC SNR, and the SNS.

First, the uncertainty of the SFR in Eq. \eqref{stellarrate} comes from the diverse data of the SFRs obtained using different methods. Ref.
\cite{Madau:2016jbv} after taking into account high redshift data ($4\leq z\leq 10$) updated the value of parameters $A$ to $D$ in Eq.  \eqref{stellarrate} to respectively
\be
A=0.01, ~B=3.2,~C=2.6,~D=6.2.\label{abcdchange}
\ee
Such large changes, particularly the decrease of the overall factor $A$ (about one third), would inevitably change (indeed decrease) the resultant DSNB spectra. We did a separate and full calculation using the updated SFR using the above new parameters. The result shows that the general shape and ordering of DSNB spectrum heights of CG models in each category are unchanged, while the overall heights  for all models are lowered by roughly a quarter and the peak positions lowered by about 2\%. What is more important is that the percentage of relative difference between heights of various CG models in each category or between different categories, and the percentage difference of total flux among different categories, are {\it quantitatively} unaffected by this change of the SFR.

\begin{figure*}[!htbp]
\vspace{-0.5cm}
\begin{center}
\includegraphics[width=0.75\textwidth]{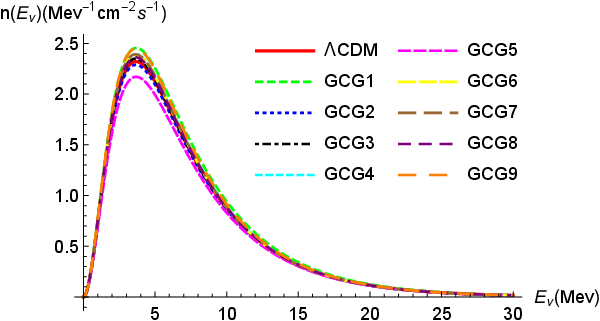}\\
(a)\\
\includegraphics[width=0.75\textwidth]{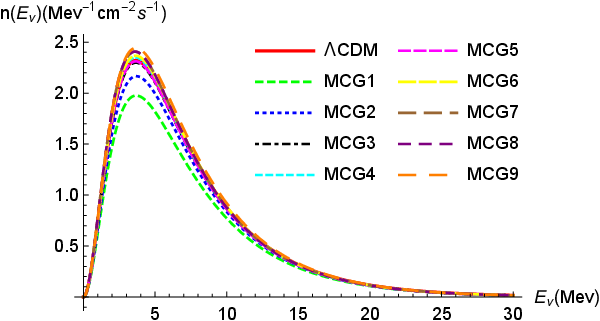}\\
(b)\\
\includegraphics[width=0.75\textwidth]{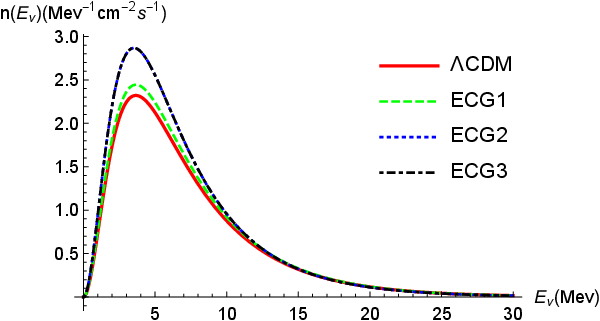}\\
(c)
\caption{The flux spectra in GCG models (a), MCG models (b) and ECG models (c) using SFR Eq. \eqref{stellarrate2} and SNR \eqref{snrate2}. The flux spectrum in $\Lambda$CDM model is drawn in each subplot using red solid line. See Table  \ref{stellarrate2} for the exact peak positions and heights. \label{figcg2}}
\end{center}
\end{figure*}

It would be more interesting to see how the DSNB spectra change when the SFR changes more dramatically, i.e., a change of the analytical form rather than only change of parameter numbers in Eq. \eqref{stellarrate}. Hopkins and Beacom used an SFR and SNR of the form
\bea
\psi(z)&=&\frac{h(a+bz)}{1+(z/c)^d} ~[\mbox{M}_\odot \mbox{year}^{-1}\mbox{Mpc}^{-3}], \label{stellarrate2}\\
R_{\textrm{SN}}&=&k^\prime \cdot \psi(z) \label{snrate2}
\eea
with $a=0.0170,~b=0.13,~c=3.3,~d=5.3,~k^\prime=0.00915~M_{\odot}^{-1}$ for Salpeter initial mass function (IMF) and  $a=0.0118,~b=0.08,~c=3.3,~d=5.2,~k^\prime=0.0132~M_{\odot}^{-1}$ for Baldry \& Glazebrook (BG) IMF \cite{Hopkins:2006bw}. The SFR with Salpeter IMF leads to an SNR with a maximum height that is 3.20 times that of the SNR in Eq. \eqref{snrate}. The location of the SNR maximum is also shifted from $z=1.86$ for Eq. \eqref{snrate} upward to $z=2.47$ for Eq. \eqref{snrate2}. The SFR with the BG IMF yields an SNR that is quite close to the one using the Salpeter IMF, with the maximum value only about 11\% lower and the peak position unchanged. Therefore we will use only the SNR obtained from SFR with Salpeter IMF in calculating the DSNB. The resultant DSNB flux spectra are shown in Fig. \ref{figcg2} for various CG models.
The peak positions and heights corresponding to these DSNB spectra are summarized in Table \ref{tb:nsfrcomb}.

Comparing Fig. \ref{figcg2} to Fig. \ref{figcg} and Table \ref{tb:nsfrcomb} to Tables \ref{tbgcg}-\ref{tbecg}, it is seen that for the new SNR, the peak heights of the DSNB fluxes are enhanced by a factor of $\sim 3.1$ for all CG models, and the peak positions are all shifted downward from roughly $E_\nu=3.8\sim 4.0$ MeV to about $E_\nu=3.5\sim 3.7$ MeV.
The enhancement of the spectra heights is easily understand because the height of the SNR \eqref{snrate2} is 3.20 times that of the SNR \eqref{snrate}. The downshift of the DSNB peak positions is
because the maximum of the new SFR is shifted to larger $z$ while the SNS in Eq.\eqref{dsnbflux} and \eqref{diffflux} exponentially suppress the large $z$ contribution. In other words, the SNR peak at larger $z$ forces the SNS to contribute more at smaller $E_\nu$ and less at larger $E_\nu$ to the DSNB, which consequently leads to the downshift of the DSNB peak positions.

Besides the above changes, we can also compare the relative difference between different CG models in the same category of or between different categories. Comparing Tables \ref{tbgcg}-\ref{tbecg} to \ref{tb:nsfrcomb}, it is seen that the relative ordering of the peak heights and positions of different models in each category are not changed. Moreover, the difference between the maximum and minimum peak heights in the GCG, MCG and ECG categories can still reach 13.1\%, 23.6\% and 17.3\% respectively. These are almost identical to the percentage difference obtained using the old SNR \eqref{snrate} and SFR \eqref{stellarrate} (13.2\%, 23.6\% and 14.9\% for the GCG, MCG and ECG categories respectively).

\begin{table*}[!htbp]
\begin{center}
\caption{The peak positions and heights with units MeV and MeV$^{-1}$cm$^{-2}$s$^{-1}$ respectively of the GCG (1-3 columns), MCG (4-6 columns) and ECG models (7-9 columns) obtained using the SFR in Eq. \eqref{stellarrate2}. \label{tb:nsfrcomb}}
\begin{tabular}{l|l|l||l|l|l||l|l|l}
  \hline\hline
Model \# & P. P. & P. H. & Model \# & P. P. & P. H. & Model \# &P. P. &P. H.\\
  \hline
GCG1 & 3.686 & 2.454 & MCG1 & 3.686 & 1.976 & ECG1 &  3.670 &  2.444 \\
GCG2 & 3.657 & 2.288 & MCG2 & 3.701 & 2.166 & ECG2 &  3.528 &  2.867 \\
GCG3 & 3.641 & 2.350 & MCG3 & 3.662 & 2.299 & ECG3 &  3.528 &  2.867 \\
GCG4 & 3.647 & 2.393 & MCG4 & 3.638 & 2.350 &  &  &  \\
GCG5 & 3.679 & 2.170 & MCG5 & 3.649 & 2.311 &  &  &  \\
GCG6 & 3.630 & 2.389 & MCG6 & 3.657 & 2.367 &  &  &  \\
GCG7 & 3.637 & 2.392 & MCG7 & 3.600 & 2.406 &  &  &  \\
GCG8 & 3.648 & 2.362 & MCG8 & 3.599 & 2.408 &  &  &  \\
GCG9 & 3.660 & 2.447 & MCG9 & 3.670 & 2.443 &  &  &  \\
  \hline\hline
\end{tabular}
\end{center}
\end{table*}

For the SNS in Eq. \eqref{snnspec}, the uncertainty of the parameters originate from the low statistics of the SN 1987A  neutrino events and the uncertainty of supernova numerical simulations. Aside from the total energy parameter $E_{\bar{\nu}_e,tot}$ whose variation will only result in an overall factor in the DSNB spectrum, the neutrino temperature $T$ is observationally constrained to 4-6 MeV \cite{Beacom:2010kk}. We changed $T$ in our calculation from 5 MeV to 4 and 6 MeV respectively. It is found that again, the overall shape and ordering of heights of the DSNB spectra for different CG models are unchanged, although the absolute heights can be changed by about one half and peak position is shifted downward or upward for about one fifth. Moreover, similar to the effect of different SFR, the difference among the DSNB spectrum heights of various CG models in each category, as well as the difference of total flux between different CG categories, are still at the same percentage.

It is instructive to see how a more dramatic change of the SNS will affect the DSNB spectra. For this purpose, we also did a full computation for an improved SNS given in the form
\cite{Keil:2002in, Tamborra:2012ac, Tamborra:2014hga}
\bea
\frac{\dd N_\nu(E_\nu^\prime)}{\dd E_\nu^\prime}&=& \frac{E_{\bar{\nu}_e,tot}}{\langle E\rangle } \frac{E^{\prime \alpha}}{\Gamma(\alpha+1)}\left(\frac{\alpha+1}{\langle E\rangle}\right)^{\alpha+1} \nonumber\\
&&\times\exp\left(-\frac{(\alpha+1)E^\prime}{\langle E\rangle}\right) ~, \label{snnspec2}
\eea
where the average energy $\langle E\rangle=3.15T$ and the spectral index $\alpha$ is chosen to be 2.7 as a representative value.
This SNS takes into various neutrino interactions in the SN explosion and therefore is more accurate \cite{Keil:2002in}. Comparing to Eq. \eqref{snnspec}, SNS in Eq. \eqref{snnspec2} only increases very slightly for $E<10$ MeV and decreases slightly for $10\mbox{MeV}<E<20\mbox{MeV}$ (maximal difference is about 8\%). Using this SNS, a new calculation found that the DSNB flux spectra are also very slightly affected, with the DSNB peak heights in all CG models increased by about 1.6\% and peak positions essentially unchanged. Therefore the percentage of relative difference between DSNB peak heights of different CG models in each category or different categories was not affected by this new SNS.

These suggest that although the uncertainties in the SNR and SNS can affect the absolute value of the DSNB spectrum and sometimes the position of the DSNB peaks, the relative variance of the DSNB heights and total flux between different CG models are independent of the these uncertainties.

\section{Discussions and Conclusion \label{secdis}}

Using a standard SNS and typical CC SNR, we obtained the DSNB flux spectra in various GCG, MCG and ECG models. It is found that generally the shape and peak position of the flux spectrum is largely determined by the SNS, SNR. Different CG models can only cause a difference of 2.3\% in the peak positions. The spectrum height however depends strongly on the Chaplygin gas model used to do the calculation. The variance in peak heights among different GCG models, MCG models and ECG models can reach respectively 13.2\%, 23.6\% and 14.9\%. This suggests a great potential for the DSNB to constraint these models. The averaged total flux for ECG models is 10.0\%-12.7\% higher than the GCG and MCG models and therefore the total DSNB flux can also be used to discriminate the Chaplygin models.

Although we consider this work mostly as a proof of concept, it is also worthy to address the uncertainties of the input parameters. It is found that change of SFR parameters in Eq. \eqref{stellarrate} to Eq. \eqref{abcdchange}, the change of its analytical form to Eq. \eqref{stellarrate2}, the change of the SNS temperature in Eq. \eqref{snnspec} or its analytical form to Eq. \eqref{snnspec2}, will only result in a change of the overall DSNB flux height or small shift of the peak positions. The general shape of the DSNB spectra, the relative ordering of both DSNB peak heights and positions, and moreover the percentage of difference between DSNB spectrum heights of various CG models in each category are not much affected.

Finally, regarding the possible extensions of the work, the following comments are in  order. Although in this work we used the Chaplygin gas models for the cosmology, it is clear from the flux spectrum formula \eqref{dsnbflux} that for other cosmological models, such as quintessence, k-essence theories, or even modified gravity cosmologies, it is straight forward to compute a similar DSNB flux spectrum and use it to constrain the corresponding cosmological model. We are currently working along this direction.

\section*{Acknowledgement}
This work is supported by the National Nature Science Foundation of China No. 11504276, 11675015, and No. 31571797. In addition, HZ is supported in part by FWO-Vlaanderen through the project G006918N, and by the Vrije Universiteit  Brussel through the Strategic Research Program ``High-Energy Physics''.  He is also an individual FWO Fellow supported by 12G3515N.





\end{document}